\documentstyle[preprint,tighten,aps,epsf,floats]{revtex}

\begin{document}

\newcommand{\gaprox}{$ {\raisebox{-.6ex}{{$\stackrel{\textstyle >}{\sim}$}}} $}
\newcommand{\saprox}{$ {\raisebox{-.6ex}{{$\stackrel{\textstyle <}{\sim}$}}} $}
\def\L{\Lambda}

\thispagestyle{empty}

%final version 11/12/98 pm 

\hfill{\small DOE/ER/40561-31-INT98}

\hfill{\small TRI-PP-98-27}

\hfill{\small KRL MAP-238}

\hfill{\small NT@UW-99-4}

\vspace*{36pt}
\begin{center}
{\large\bf The Three-Boson System with Short-Range Interactions}

\vspace*{42pt}

{\bf P.F. Bedaque$^a$}, {\bf H.-W. Hammer$^b$}, and {\bf U. van Kolck$^{c,d}$}

\vspace*{12pt}

{\sl $^a$Institute for Nuclear Theory}\\
{\sl University of Washington, Seattle, WA 98195, USA}\\
{\tt bedaque@mocha.phys.washington.edu}\\
\vspace{10pt}
{\sl $^b$TRIUMF, 4004 Wesbrook Mall }\\
{\sl Vancouver, BC, V6T 2A3, Canada}\\
{\tt hammer@triumf.ca}\\
\vspace{10pt}
{\sl $^c$ Kellogg Radiation Laboratory, 106-38}\\
{\sl  California Institute of Technology, Pasadena, CA 91125, USA}\\
{\tt vankolck@krl.caltech.edu}\\
\vspace{10pt}
{\sl $^d$ Department of Physics}\\
{\sl University of Washington, Seattle, WA 98195, USA}

\vspace{12pt}
\end{center}

\begin{abstract}
We discuss renormalization of the 
non-relativistic three-body problem with short-range forces. 
The problem is non-perturbative at momenta of the
order of the inverse of the 
two-body scattering length. An infinite number of 
graphs must be summed, which
leads to a cutoff dependence
that does not appear in any order in perturbation theory. We argue 
that this cutoff dependence can be absorbed in one local three-body force
counterterm and compute the running of the three-body force with the
cutoff. This allows a calculation of the scattering of  a particle
and the two-particle bound state
if the corresponding 
scattering length is used as input.
We also obtain a model-independent relation between 
binding energy of a shallow three-body bound state
and this 
scattering length.
We comment on the power counting 
that organizes higher-order corrections and
on relevance of this result for the
effective field theory program in nuclear and molecular physics.
\end{abstract}

\vfill
\newpage
\setcounter{page}{1}

\section{Introduction}

The three-body system provides non-trivial testground for ideas developed
in two-body dynamics, and as such has a long and venerable history.
It is in general of very difficult solution, but
when all particles have momenta much smaller than the inverse of the 
range of interactions
it simplifies considerably while still retaining some very rich
universal features \cite{thomas,efimov,phillips}.
At such low energies, the two-body system can be attacked with many
different techniques: effective range expansion, boundary conditions
at the origin, {\it etc.}; a particularly convenient method
for the extension to a many-body context is 
that of effective field theory (EFT) \cite{gospel}.
Here we use EFT to solve the
three-body system with short-range interactions 
in a systematic momentum expansion.

Generically the sizes of bound states
made out of $A$ particles of mass $m$
are all comparable to the range $R$ of interactions.
Similarly, the dimensionful scattering parameters 
--- the two-body scattering length $a_2$, 
the two-body effective range parameter $r_2$, and so on---
are also of the same order, $R\sim a_2 \sim r_2\sim \ldots$ 
The $A$-body amplitudes reduce at low energy $k^2/m$
to perturbative expansions in $k R$ \cite{braaten}. 
Less trivial is the case where the interactions are fine-tuned so that
the two-body system has a shallow (real or virtual) bound state 
---that is, a bound state
of size $\sim a_2$ much larger than the range of the interactions
$R\sim r_2$. This is a case of interest in nuclear physics,
where the deuteron is large compared to the Compton wavelength
of the pion $1/m_\pi$, and in molecular physics,
where shallow molecules such as the $^4$He ``dimer'' 
can be over an order of magnitude larger than 
the range of the interatomic potential.

In this fine-tuning scenario,
the $A$-body system still can be treated in a perturbative
expansion in $k a_2$ in the scattering region where all momenta
are of order $k\ll 1/a_2$. 
Bound states, however, correspond to $k \gaprox  a_2$ and
demand that a certain class of interactions be 
iterated an infinite number of times. 
In the two-body system, it can be
shown \cite{1stooge,3musketeers} that one needs to sum exactly only 
two-body contact interactions which are momentum independent.
This resummation generates a new expansion in powers of $k R$ where
the full dependence in $k a_2$ is kept.
Contact interactions with increasing number of derivatives
can be treated as perturbative insertions of increasing
order.
First and second corrections, for example, 
are determined by the
insertion of a two-derivative contact interaction,
which encodes information about the effective range;
to these first three orders, all there is is $S$-wave scattering
with an amplitude equivalent (up to higher-order terms) to
the first two terms in an effective range expansion.
The EFT for the two-particle systems is thus equivalent to 
effective range theory 
\cite{1stooge,3musketeers,Gegeliaetal}; it
is valid even at $k\sim 1/a_2$ and,
in particular, for bound states of size $\sim a_2$.

There has been
great progress recently in dealing with this problem in the two-body case
\cite{edict}.
Ultraviolet divergences appear in graphs with leading-order interactions
and their resummation contains
arbitrarily high powers of the cutoff.
A crucial point is that
this cutoff dependence
can be absorbed in the coefficients of the leading-order
interactions themselves.  All our ignorance
about the influence of short-distance physics on low-energy phenomena
is then embodied in these few coefficients,
and EFT has predictive power.

The question we want to answer in this paper is whether
the leading two-body interactions are sufficient to approximately 
describe the three-body system in the same energy
range, or whether we also need to include three-body interactions in leading 
order. The answer hinges on the relative size of three-body interactions,
so this question is intimately related to the renormalization
group flow of three-body interactions with the mass scale
introduced in the regularization procedure. 
This flow, in turn, depends on
the behavior of the sum of two-body contact contributions
to three-body amplitude as function of the renormalization scale,
or equivalently, as function of the ultraviolet cutoff $\Lambda$. 
What makes this problem different from standard field theory examples
is that it does 
{\it not} have a
perturbative expansion in a small coupling constant, and thus from
the start involves an infinite number of diagrams.  

The extension of the EFT program
to three-particle systems in fact presents us with a puzzle \cite{1stoogetoo}.
The two-nucleon ($NN$) system has a shallow real bound state (the deuteron)
in the $^3S_1$ channel
(and a shallower virtual bound state in the $^1S_0$ channel).
Information about the three-nucleon system is accessible
in nucleon-deuteron ($Nd$) scattering, which proceeds
via two channels of total spin $J=3/2$ and $J=1/2$. 
Because of the Pauli principle we expect smaller three-body
contact forces in the quartet channel than in the doublet channel.
Assuming three-body forces to have a size completely determined
by $R$ according to naive dimensional analysis,
the $J=3/2$ $Nd$ amplitude only receives contributions of
three-body forces at relative $O((kR)^6)$,
so up to (and including) relative $O((kR)^2)$
the $J=3/2$ $Nd$ amplitude depends on the $NN$ amplitude only
\cite{2stooges,3stooges}.
At low energies, the $J=3/2$ $Nd$ amplitude can be obtained by
solving a single integral equation. 
We have shown that it is ultraviolet convergent,
and its low-energy end independent of the cutoff;
moreover, with parameters
entirely determined by $NN$ data, 
it predicts low-energy phase shifts in excellent agreement
with the experimental scattering length \cite{2stooges}
and with an existing phase-shift analysis \cite{3stooges}. 
The $J=1/2$ $Nd$ amplitude constructed out of the
$NN$ amplitude to the same order can be obtained
from a pair of coupled integral equations.
However, despite describing a sum of
ultraviolet finite diagrams, numerical experimentation shows that it
does {\it not} converge as the cutoff is increased.
A system of three bosons exhibits a similar problem in the easier
context of a single integral equation. 
For simplicity, we limit ourselves here to the latter case,
which is of relevance to $^4$He molecular systems.
Similar arguments but more cumbersome formulas apply 
to the $J=1/2$ three-fermion amplitude, which we will
address in a later publication. 

Our study concerns momenta such that all forces can be considered
short-ranged, but 
it is also relevant to higher momenta where we start
to resolve the short-range dynamics. In general, what we have been
considering as short-range dynamics
will itself have structure and consist of longer- and shorter-range
components. 
The latter will still be described
by contact interactions in the new EFT appropriated to these higher
energies.
Elements of the discussion presented here will 
permeate this more complicated renormalization scenario.
In the nuclear case, as we increase energy
we start seeing effects of pion propagation,
and interactions have both pion-exchange and contact components
of similar sizes \cite{wein}.
It has recently been argued \cite{3musketeers}
that, because of fine-tuning, at moderate
energies contact interactions
are actually larger than pion effects by a factor of 2 or 3, 
so that pion interactions
can be treated as corrections.
A number of applications show that
deuteron physics can be fruitfully dealt with this way
\cite{ocordaumdospuxasacos...}.
The momentum where this power counting breaks down
is currently matter of controversy \cite{controversy}.  
Our results hold unchanged throughout the region of validity 
of this power counting.

After setting up our formalism in Sect. \ref{sect2},
we show in Sect. \ref{sect3} that in general
the sum of two-body contact contributions
to the particle/bound-state scattering
amplitude, while finite, does not converge
as the cutoff $\Lambda$ is increased. 
We then present evidence in Sect. \ref{sect4} that 
the leading-order cutoff dependence can be absorbed in a 
local three-body force
with one single parameter $\Lambda_\star$.
In particular, we derive an approximate analytic formula for
the dependence of the bare three-body on the cutoff  $\Lambda$.
As a consequence, although input
from a three-body datum
is necessary, 
the EFT retains its predictive power.
For example, if the 
three-body scattering length $a_3$
is used
to determine $\Lambda_\star$, then
the amplitude at (small) non-zero energies can be predicted
and is cutoff independent. 
If $\Lambda_\star$ is such that there exists 
a (ground or excited) bound state large enough to be 
within the range of the EFT expansion, its binding energy $B_3$
can be predicted as well. 
As an example, we 
consider the $^4$He trimer.
In any such physical system, $\Lambda_\star$
is determined by the dynamics of  
the underlying theory as a certain (possibly complicated)
function of its parameters.
Different models for the underlying dynamics that are fit to 
the same two-body scattering
data will correspond to
different values for $a_3$ and $B_3$, and in principle
could cover the whole plane $B_3 \times a_3$.
The EFT, however, predicts that in leading order
these models can be distinguished by only one number, $\Lambda_\star$,
and therefore that they would lie on a {\it curve}  $B_3= B_3(a_3)$.
Indeed, in a nuclear context this has been observed,
and is called the Phillips line \cite{phillips}.
We discuss bound states
and construct the corresponding bosonic line in Sect. \ref{sect5}.
In Sect. \ref{sect6}
we discuss how corrections to this leading order can be handled
in a systematic fashion.
In Sect. \ref{sect7} we offer conclusions on our 
extension of the EFT program to three-particle systems with
large two-body scattering lengths.
Some of the results presented here
were briefly reported in Ref. \cite{stoogeletter}.

\section{Lagrangian and Integral Equation} \label{sect2}

Particles with momenta much smaller than their mass $m$ propagate
as non-relativistic particles. If momenta are also small compared
to the range of interaction $R$, the (bare) interaction 
can be approximated by a sequence
of contact interactions, with an increasing number of derivatives.
This is true regardless of the fine details of the interaction;
information about these details is encoded in the actual values
of the coefficients of the contact interactions.
The most general Lagrangian involving a non-relativistic
boson $\psi$ and invariant under  
small-velocity Lorentz, parity, and time-reversal transformations
is 
\begin{equation}
\label{lag}
{\cal L}  =  \psi^\dagger
             (i\partial_{0}+\frac{\vec{\nabla}^{2}}{2m})\psi
 - \frac{C_0}{2} (\psi^\dagger \psi)^2
 - \frac{D_0}{6} (\psi^\dagger\psi)^3 + \ldots ,\nonumber
\end{equation}
where the ellipsis stand for terms with more derivatives and/or fields;
those with more fields will not contribute to
three-body amplitudes,
while those with more derivatives are suppressed at low
momenta.

The scope of this EFT in the two-body sector is by now well understood
\cite{edict}. 
The two-body amplitude will contain, in addition to
``analytic'' terms from two-particle contact interactions, 
also their iteration, which produces loops
whose non-analytic terms are responsible for the unitarity cut.
In leading order, 
one has to consider only the $C_0$ term iterated
to all orders, which is equivalent to the effective range
expansion truncated at the level of the scattering length.
The ultraviolet divergences found can all be absorbed in the parameter
$C_0$. The renormalized value of this parameter is
$C_0= 4\pi a_2/m$. For $a_2>0$ there is a real 
$S$-wave bound state at energy
$-B_2=-1/m a_2^2$.
It can be further shown that the renormalized on-shell two-body amplitude 
forms an expansion in momenta that breaks down only at
momenta of $O(1/R)$ \cite{1stooge}. In the fine-tuning
scenario we are interested in, for momenta of magnitude
typical of the shallow $S$-wave bound state, $k \sim 1/a_2$,
the expansion is in the small parameter $R/a_2\ll 1$. 
We want here to extend this approach to the three-body system and
establish the order in the EFT expansion in which
a three-body force first appears.

It is convenient \cite{transvestite}
to rewrite this theory
by introducing a dummy field $T$ with quantum
numbers of two bosons, referred to from here on as ``dimeron''
(in analogy to dibaryon in the nuclear case).
It is straightforward to show 
---for example, by a Gaussian path integration---
that the Lagrangian Eq. (\ref{lag}) is equivalent to
\begin{equation}
\label{lagt}
{\cal L}  = \psi^\dagger
             (i\partial_{0}+\frac{\vec{\nabla}^{2}}{2m})\psi
         + \Delta T^\dagger T
  -\frac{g}{\sqrt{2}} (T^\dagger \psi\psi +\mbox{h.c.})\nonumber\\
      +h T^\dagger T \psi^\dagger\psi
 +\ldots
\end{equation}
\noindent
The arbitrary scale $\Delta$ is included in (\ref{lagt}) only to give
the field $T$ the usual mass dimension of a heavy field.
Observables will depend on
the parameters appearing explicitly
in Eq. (\ref{lagt}) only through the combinations
$g^2/\Delta\equiv C_0$
and $-3hg^2/\Delta^2\equiv D_0$.

Two-particle contact interactions are thus replaced by  
the $s$-channel propagation of the
dimeron field; the two-body amplitude
is completely determined by the full dimeron self-energy,
that is, by the dimeron propagator dressed by particle loops.
Three-particle contact interactions are likewise rewritten as
particle-dimeron contact interactions, and so on.

Let us consider elastic particle/bound-state scattering. 
(Three-body bound states manifest themselves as poles
at negative energy.
Both the inelastic channel and three-nucleon scattering 
involve the same type of diagrams and can be obtained by
a  straightforward extension of the following arguments.)
We restrict ourselves to external
momenta $Q\sim 1/a_2$ and energies $Q^2/m\sim 1/ma_2^2$.
In leading order in a low-momentum expansion,
the (bare) dimeron propagator is simply a constant $i/\Delta$,
while the propagator 
for a particle of four-momentum $p$ reduces to the usual non-relativistic
propagator 
\begin{equation}
iS(p) = {i\over  p^0- \frac{\vec{p}^{\,2}}{2m} +i\epsilon}, \label{Nprop}
\end{equation}
with $S=O(m/Q^2)$.

\begin{figure}[t] %bh]
\begin{center}
\epsfxsize=12cm
\centerline{\epsffile{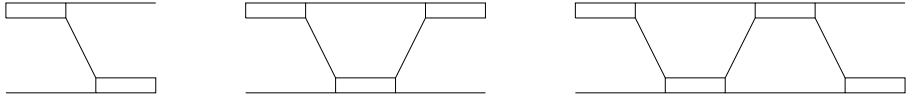}}
\end{center}
\caption{Bare pinball diagrams up to $O(g^6)$. 
A single (double) line represents a
particle (bare dimeron) propagator.}
\label{fig1}
\end{figure}

A few of the first (connected) 
diagrams in perturbation theory in the coupling constant $g$
are illustrated in Fig. \ref{fig1}. 
In second order in the coupling, there is a single, tree diagram,
of $O(m g^2/Q^2)$. 
The second diagram is of fourth order and has one loop; it is 
power-counting convergent and of 
$O((m g^2)^2/4\pi Q \Delta)$.
The third diagram is of sixth order and has two-loops;
it is logarithmically divergent in the ultraviolet 
cutoff $\Lambda$ and of 
$O((m g^2)^3\ln (\Lambda/\Delta)/$ $(4\pi \Delta)^2 )$.
It is clear that increasing the order brings factors of 
$m g^2 Q/4\pi \Delta$,
so the divergences get progressively worse.
At this point, one might jump to the conclusion 
that removal of divergences would require an
increasing number of three-body force counterterms, starting with 
$h\sim ((m g^2)^3/(4\pi \Delta)^2) \ln (\Lambda/\Delta)$.

\begin{figure}[b]
\begin{center}
\epsfxsize=12cm
\centerline{\epsffile{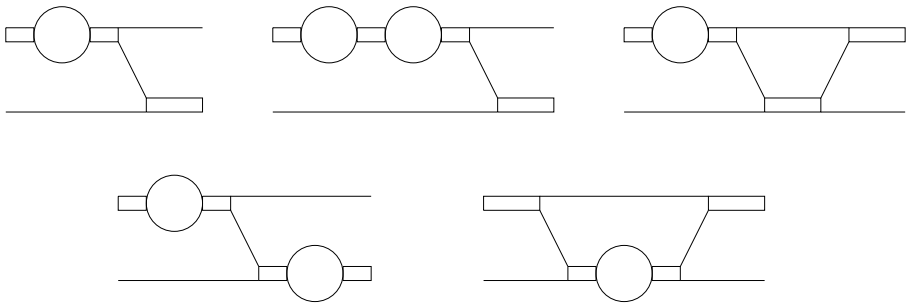}}
\end{center}
\caption{Pinball diagrams with bubbles up to $O(g^6)$.}
\label{fig2}
\end{figure}

However, there are other diagrams that appear in the same orders:
besides the bare ``pinball'' diagrams of Fig. \ref{fig1}, we also
have those which contain insertions of nucleon loops
in dimeron propagators, the first few of which are shown in 
Fig. \ref{fig2}.  
All but one of the diagrams 
of fourth and sixth orders,
contribute to (off-shell) wave-function renormalization.
The remaining sixth-order diagram 
is of the form of the fourth-order bare pinball. 
The extra loop contributes a linearly divergent integral.
The divergent piece $\propto \Lambda$ can be absorbed in $g^2/\Delta$,
and for cutoffs $\Lambda\sim 1/R$ the convergent
terms that go $\propto Q^2/\Lambda$ and higher inverse
powers of $\Lambda$ are of the size of
other already disregarded higher-order terms.
This is just two-body renormalization at work in a three-body context:
the net effect of a nucleon loop in the dimeron
propagator is the remaining, non-analytic term 
of $O(m g^2 Q/4 \pi \Delta)$.

If we were interested in momenta much smaller than the typical 
bound-state momentum, this would be essentially the end of the story.
Things are more interesting for  $Q\sim 1/a_2$,
in which case the coupling constant expansion is an expansion in
$m g^2 Q/4 \pi \Delta \sim 1$.
In the two-body subsystem, all graphs built up from leading
interactions are equally important. Whenever a bare dimeron
propagator appears, so do all bubble insertions in it.
It is not difficult to show \cite{1stooge} that the two-body problem can 
still be renormalized by absorbing in $g^2/\Delta$ the factors
of $\Lambda$ from more insertions of nucleon loops.
(From here on  $g^2/\Delta$ refers to the sum of 
bare and loop contributions.) 
But what happens with the divergences in the graphs of Fig. \ref{fig1}?

\begin{figure}[tb]
\begin{center}
\epsfxsize=13cm
\centerline{\epsffile{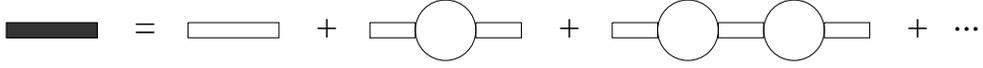}}
\end{center}
\caption{Dressed dimeron propagator.}
\label{fig3}
\end{figure}

The three-body amplitude can now be reexpressed in terms of
the graphs in Fig. \ref{fig1} but with
the dressed dimeron propagator shown in Fig. \ref{fig3}, 
\begin{equation}
i \Delta(p) =  \frac {-i}{- \Delta
             + \frac{m g^{2}}{4\pi}
               \sqrt{-m p_0+\frac{\vec{p}^2}{4}-i\epsilon} +i\epsilon},
                                   \label{Dprop}
\end{equation}
\noindent
substituted for the bare propagator.  
The extra $\sqrt{}$ in the denominator improves ultraviolet 
convergence: now pinball loops carry factors of 
$m g^2 Q/4\pi (\Delta+Q)$. 
All diagrams go as $1/\Lambda^2$ for $Q\sim \Lambda$,
{\it i.e.}, are power counting finite.
The resummation into a dressed propagator
accounts for cancellations among divergences found in diagrams with
bare propagators. 
One could expect, based on perturbation theory experience, that
for $\Lambda \gg 1/a_2$ the low-energy end of this three-body amplitude 
is to a good approximation cutoff independent, 
and that the three-body amplitude converges as $\Lambda\rightarrow \infty$.

\begin{figure}[t] %bh]
\begin{center}
\epsfxsize=15cm
\centerline{\epsffile{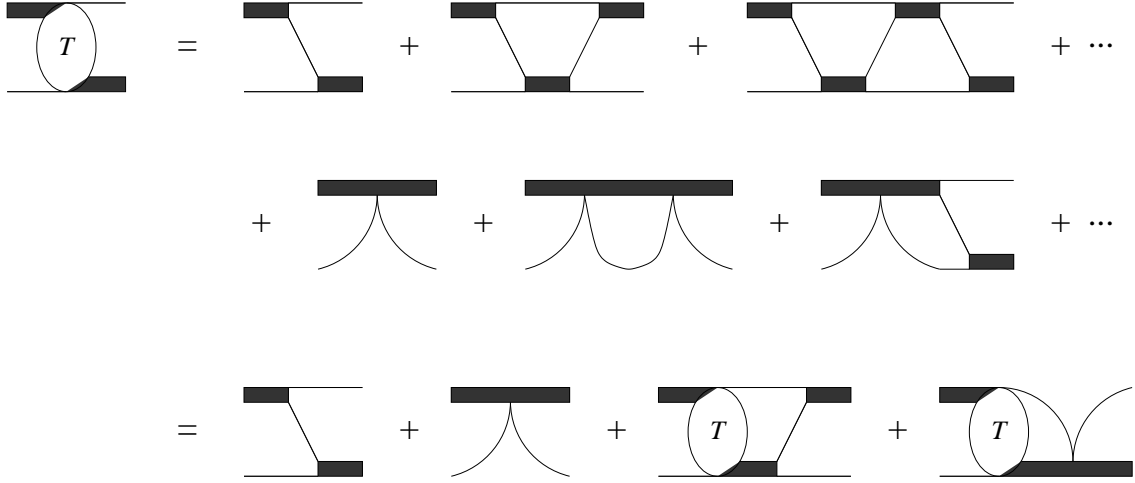}}
\end{center}
\caption{The amplitude $T$ for particle/bound-state scattering
as a sum of dressed pinball and three-body-force diagrams (first 
and second lines)
and as an integral equation (third line).}
\label{fig4}
\end{figure}

Unfortunately, things are not so straightforward. 
Since all diagrams are of the same order, $O(m g^2 Q/4\pi \Delta) \sim 1$,
the three-body amplitude is actually the solution of an integral equation.
This equation, including the three-body force, is depicted in 
Fig. \ref{fig4}. We choose the following kinematics: the incoming 
particle and bound state
are on-shell with four-momenta $(k^2/2m, -\vec{k})$ and 
$(k^2/4m-B_2, \vec{k})$,  
respectively.
The outgoing particle and bound state
are off-shell with four-momenta 
$(k^2/2m-\varepsilon, -\vec{p})$ and 
$(k^2/4m-B_2+\varepsilon, \vec{p})$; the on-shell point has 
$\varepsilon=k^2/2m - p^2/2m$ and $p=k$.
The total energy is  $E = 3k^2/4m - B_2$.
Denoting the blob in Fig. \ref{fig4} with this kinematics by 
$it(\vec{k},\vec{p},\varepsilon)$, we have
\begin{eqnarray}
it(\vec{k},\vec{p},\varepsilon) & = &
     -2 g^2 iS(-k^2/4m -B_2 +\varepsilon,\vec{p}+\vec{k})
         + ih\nonumber\\
   & &\! +\lambda \! \int \! \! \frac{d^4q}{(2\pi)^4}  
                  iS(k^2/2m -\varepsilon-q_0,-\vec{q}) \!
\left[-2 g^2 iS(-k^2/4m -B_2 +2\varepsilon+q_0,\vec{p}+\vec{q})
    + ih\right] \nonumber\\
& & \; \; \; \; \; \; \; \; \; \; \; \; \; \; \; \;
    i\Delta(k^2/4m -B_2 +\varepsilon+q_0,\vec{q}) \,
                   it(\vec{k},\vec{q},\varepsilon+q_0).
\end{eqnarray}
\noindent
Here $\lambda=1$ for the bosonic case we are considering.
The same equation is valid
in the case of fermions, with different values of $\lambda$. In particular, 
for three nucleons in a spin $J=3/2$ state under two-body interactions
only, this equation holds
with $\lambda=-1/2$ and $h=0$.
(For three nucleons in a spin $J=1/2$ state
a pair of coupled integral equations results, with properties
similar to the bosonic case.)

After performing the $q_0$ integration
we can set $\epsilon=k^2/2m - p^2/2m$ and, defining  
$t(\vec{k},\vec{p})\equiv t(\vec{k},\vec{p},k^2/2m - p^2/2m)$,
we have
\begin{eqnarray}
t(\vec{k},\vec{p})& = &\frac{2 m g^2}{\frac{k^2}{4}+p^2+mB_2 
                                      +\vec{p}\cdot\vec{k}}
   +h \nonumber \\ 
     & & +8\pi\lambda 
                   \int \!  \frac{d^3q}{(2\pi)^3}
                   \frac{t(\vec{k},\vec{q})}{-\frac{1}{a_2}
                         +\sqrt{\frac{3q^2}{4}-mE } }
        \left[\frac{1}{-\frac{3 k^2}{4} +mB_2 +q^2 +p^2+\vec{p}\cdot\vec{q}}
                   +\frac{h}{2 m g^2}
             \right]. 
\end{eqnarray}
\noindent
Projection on 
the $S$-wave is obtained by integration over the angle between
$\vec{p}$ and $\vec{k}$, resulting in
\begin{eqnarray}
t(k,p)&=&\frac{m g^2}{p k} 
         \ln \left( \frac{p^2+pk+k^2-mE}{p^2-pk+k^2-mE} \right) + h\nonumber \\
    & &+\frac{2\lambda}{\pi}\int_0^\Lambda dq \frac{t(k,q) q^2}{-\frac{1}{a_2}
              +\sqrt{\frac{3q^2}{4}-mE } }
               \left[ \frac{1}{pq}\ln \left( \frac{p^2+pq+q^2-mE}
                                                  {p^2-pq+q^2-mE} \right)
                           +\frac{h}{m g^2} \right].
\end{eqnarray}

The scattering amplitude $T(k)$ is given by
\begin{equation}
T(k)= \sqrt{Z} t(k,k) \sqrt{Z},
\end{equation}
\noindent
where $Z$ is given by
\begin{equation}
Z^{-1}=i \left.\frac{\partial}{\partial p_0} 
              (i\Delta(P))^{-1} \right|_{p_0=-B_2}=
             \frac{m^2g^2}{8\pi\sqrt{mB_2}}.
\end{equation}
\noindent
It is customary to define the function $a(k,p)$, 
\begin{equation}
\frac{a(k,p)}{p^2-k^2}=\frac{m a_2}{8 \pi}\frac{Zt(k,p)}{-\frac{1}{a_2}
+\sqrt{\frac{3p^2}{4}-mE }},
\end{equation}
\noindent
that on shell reduces to
\begin{equation}
a(k,k)=\frac{m}{3\pi}T(k).
\end{equation}
\noindent
In particular, $a(0,0)=-a_3$. The equation satisfied by $a(k,p)$ is
\begin{equation}
a(k,p)
=M(k,p;k)+\frac{2\lambda}{\pi}\int_0^\Lambda dq\ M(q,p;k)
\frac{q^2}{q^2-k^2-i\epsilon} a(k,q), \label{aeq}
\end{equation}
\noindent
with the kernel
\begin{equation}
M(q,p; k)= \frac{4}{3} 
     \left(\frac{1}{a_2}+\sqrt{\frac{3p^2}{4}-mE}\right)
   \left[\frac{1}{pq}{\rm ln}
    \left(\frac{q^2+q p +p^2-mE}
               {q^2-q p +p^2-mE}\right)
    +\frac{h}{mg^2} \right]. \label{kernel} 
\end{equation}
\noindent
Eqs. (\ref{aeq}) and (\ref{kernel}) reduce to the expressions found in
Refs. \cite{skorny,1stoogetoo,2stooges,3stooges}
when $h=0$. 
Note that the perturbative series shown in Fig. \ref{fig4} corresponds
to a perturbative solution of the integral equation
for small $\lambda$. 

The solution of Eq. (\ref{aeq}) is complex even below the threshold for the
breakup of the two-particle bound state due to the $i\epsilon$ prescription.
To facilitate our discussion we will use below the function
$K(k,p)$ that satisfies the same Eq. (\ref{aeq}) as $a(k,p)$ but with the
$i\epsilon$ substituted by a principal value prescription. $K(k,p)$
is real below the breakup threshold. $a(k,k)$ and, consequently, 
the scattering matrix can be obtained from  $K(k,p)$ through
\begin{equation}
a(k,k)=\frac{K(k,k)}{1-i k K(k,k)}.
\end{equation}

\section{The Problem} \label{sect3}

In order to understand the ultraviolet behavior of the theory,
let us first take $h=0$.
When $p\gg 1/a_2$ (but $k\sim 1/a_2$), the inhomogeneous term
is small ($O(1/pa_2)$), the main contribution to the integral comes from
the region $q\sim p$, and the amplitude satisfies the 
approximate equation
\begin{equation}
K(k,p)= \frac{4\lambda}{\sqrt{3}\pi}
                   \int_0^\L \: \frac{dq}{q}\ 
                   {\rm ln} \left(\frac{q^2+pq+p^2}{q^2-pq+p^2}\right)
                   K(k,q).\label{asphieq} 
\end{equation}

Now, in the limit $\Lambda\rightarrow \infty$ there is no scale left.
Scale invariance suggests solutions of the form  
$K(k,p)= p^s$, which exist only if
$s$ satisfies 
\begin{equation}
1- \frac{8\lambda}{\sqrt{3}s} 
   \frac{\sin\frac{\pi s}{6}}{\cos\frac{\pi s}{2}}=0.
\label{rhoeq}
\end{equation}
\noindent
If $K(k,p)$ is
a solution, $K(k,p_\star^2/p)$ 
is also a solution for arbitrary
$p_\star$. Because of this  additional
symmetry, the solutions of Eq. (\ref{rhoeq}) come in pairs. 

For $\lambda<\lambda_c={3 \sqrt{3}\over 4 \pi}\simeq 0.4135$, Eq.
(\ref{rhoeq}) has only real roots.
For example, if $\lambda=-1/2$, 
there are solutions $s=\pm 2, \pm 2.17, \ldots$ 
In this case
an acceptable solution of Eq. (\ref{asphieq})
decreases in the ultraviolet, $K(k,p\gg 1/a_2)= C p^{-|s|}$.
For finite cutoff, the solution should still have this form 
as long as $p\ll \Lambda$. 
The overall constant $C=C(\Lambda)$ 
cannot be fixed
from the homogeneous asymptotic equation, but is determined
by matching the asymptotic solution onto
the solution at $p\sim 1/a_2$ of the full Eq. (\ref{aeq}).  
Because of the fast ultraviolet convergence, the full solution to
Eq. (\ref{aeq}) is expected to be insensitive to the choice
of regulator, so that a well-defined
$\lim_{\Lambda\rightarrow \infty} C(\L)$ can be found. 
This behavior can indeed be seen in
a numerical solution of Eq. (\ref{aeq}) \cite{2stooges,3stooges}.

For $\lambda=1$, on the other hand, there
are purely imaginary solutions: $s=\pm is_0$, where $s_0 \simeq 1.0064$.
The solution of Eq. (\ref{asphieq}) as $\Lambda\rightarrow \infty$
is 
\begin{equation}
K(k,p\gg 1/a_2)= C \: 
        \cos \left(s_0{\rm ln} \frac{p}{p_\star}\right). \label{assol}
\end{equation}
\noindent
Again, for finite cutoff this should hold at least for $p\ll \Lambda$,
with $p_\star=p_\star(\L)$ determined by the cutoff.
On dimensional grounds,
\begin{equation}
 p_\star(\Lambda)=\exp(- \delta/s_0)\: \Lambda, \label{xstar}
\end{equation}
\noindent
where $\delta$ is a dimensionless, cutoff independent number. 

There are now {\it two} constants 
$C(\Lambda)$ and $\delta$:
the solution of Eq. (\ref{aeq}) is not 
unique in the limit $\Lambda\rightarrow \infty$ \cite{danlebed}.
The undetermined phase $p_\star$ arises from
the  symmetry $K(k,p)\rightarrow K(k,p_\star^2/p)$.
When $\lambda=-1/2$ we can dismiss the solution that
grows in the ultraviolet since the integral in the Eq. (\ref{aeq}) will not
converge, but 
in the $\lambda=1$ case there is no way to select a preferred
oscillatory solution. 
The same problem exists for any $\lambda$ that yields purely imaginary
solutions for $s$ in Eq. (\ref{rhoeq}). 
From now on, for definiteness we specialize to the case 
of purely imaginary $s$.

\begin{figure}[bt]
\begin{center}
\epsfxsize=8cm
\centerline{\epsffile{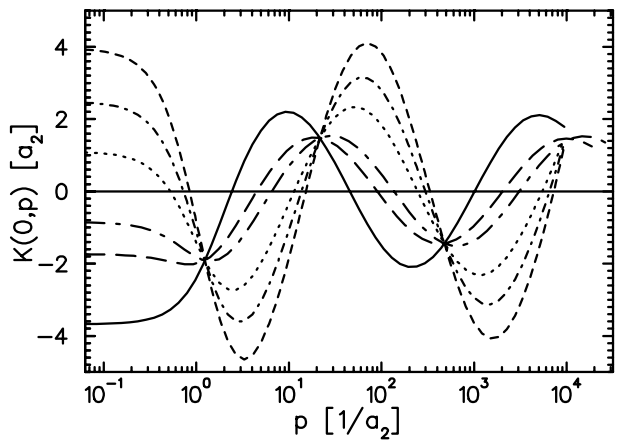}}
\end{center}
\caption{Amplitude $K(0,p)$ as function of the momentum $p$. 
Full, dashed, and dash-dotted curves are
for $H=0$ and $\Lambda=1.0,\,2.0,\,3.0\times 10^4 a_2^{-1}$, respectively.
Dotted, short-dash-dotted, and short-dashed curves are for
$\Lambda=1.0 \times 10^4 a_2^{-1}$ and $H=-6.0,\,-2.5,\,-1.8$, respectively.}
\label{fig5}
\end{figure}

Solutions $K(0,p)$ of Eq. (\ref{aeq}) for $\lambda=1$ and
finite $\Lambda$ (but $h=0$) are shown in Fig. \ref{fig5}. 
We have verified that in the region $1/a_2\ll p\ll \Lambda$ 
the solutions indeed
have the form (\ref{assol}), with $s_0 = 1.02\pm 0.01$. 
The phase $p_\star$ was found approximately linear in the cutoff
in accordance with Eq. (\ref{xstar}):
$\delta= 0.76\pm 0.01$ is cutoff independent.
We have also found that the amplitude is  
\begin{equation}
C= -\frac{\gamma}{\cos \left(s_0 \ln (p_\star a_2) +\epsilon\right)} 
\label{Ceq}
\end{equation}
\noindent
with $\gamma=1.50\pm 0.03$ and $\epsilon\sim 0.08$,
both cutoff independent.
This  generates the striking meeting points at 
$p_n=a_2^{-1} \exp([(n+1)\pi-\epsilon]/s_0)
\sim 0.92 a_2^{-1} (22.7)^{(n+1)}$, $n$ an integer, 
that can be seen in  Fig. \ref{fig5}.

Apart from these meeting points, the amplitude does not
show signs of converging as $\L\rightarrow \infty$.
We are forced to conclude that {\it terms that are $O(Q/\Lambda)$ 
cannot be dropped}, as it is usually done in 
calculations in finite orders in perturbation theory.
The presence of these terms determines a unique phase in the 
asymptotic region. 
The solution for small $p$ is to be matched at an intermediate scale
to the large-$p$ solution, so the cutoff dependence
leaks into the small-$p$ region. 
Small differences in
the asymptotic phase lead to large differences in,
for example, the particle/bound-state
scattering length.

\begin{figure}[tb]
\begin{center}
\epsfxsize=8cm
\centerline{\epsffile{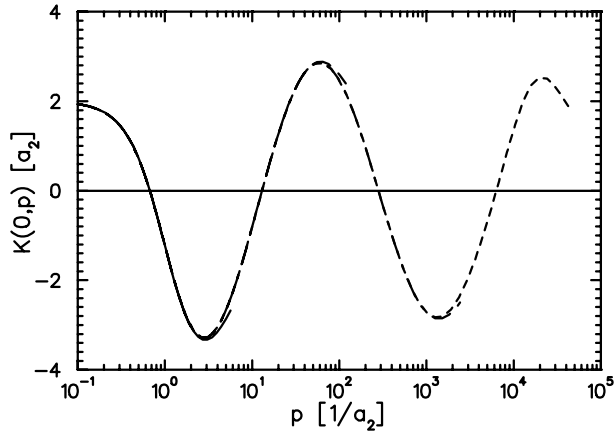}}
\end{center}
\caption{Amplitude $K(0,p)$ for $\Lambda_{0, 1, 2, 3}$,
in the family of $\Lambda_0=5.72 a_2^{-1}$,
corresponding to $H(\Lambda_n)=0$, for $a_3=-2.0 a_2$.}
\label{fig6}
\end{figure}

Since $K(k,p\gg 1/a_2)$ is given by Eqs. (\ref{assol})--(\ref{Ceq}),
it is the same within a discrete family of cutoffs 
$\Lambda_n= \Lambda_0 \exp(n\pi/s_0)\simeq \Lambda_0 (22.7)^n$, 
$n$ an integer. 
We expect that the low-energy solution will then be
invariant for cutoffs in this family. 
If $\Lambda_0$ is fixed in such a way that  the three-body
scattering length $a_3$ is reproduced,
then the same $a_3$ should result from any of the $\Lambda_n$'s
in the same family.
Moreover, the amplitude $K(k,p)$ around 
$p=k=0$ should be similar as well.
This is indeed what we find:
in Fig. \ref{fig6}, we plot the low-energy solutions for
cutoffs in the family of 
$\Lambda_0=5.72 a_2^{-1} $ ($\Lambda_1=122.2 a_2^{-1}\simeq 22.7 \Lambda_0$, 
{\it etc.}) fitted to give $a_3=-2.0 a_2$ (cf. Fig. \ref{fig9}).
For the appropriate $a_3/a_2$, we find the same cutoffs as in 
Ref. \cite{kharchenko}.
This generalizes the results of Ref. \cite{kharchenko}:
we see that a family of cutoffs determines not
only the scattering length
but also the (low-)energy dependence of the amplitude.
As we vary families ({\it i.e.}, $\Lambda_0$),
however, the low-energy
behavior of the amplitude shows strong cutoff dependence.
This leakage of high-momentum behavior into the low-momentum physics 
is indication that we are not performing renormalization
consistently with our expansion.

Note that if one were to truncate the series
of diagrams in Fig. \ref{fig4} at some finite number of loops
one would miss 
the asymptotic behavior of $K(k,p)$ that
generates this cutoff dependence.
This is because $s_0$ (and its expansion in powers of $\lambda$)
vanish in a neighborhood of $\lambda=0$.
The truncation of the series in  Fig. \ref{fig4}
is equivalent to perturbation theory in $\lambda$,
and cannot produce a non-vanishing $s_0$.
We are here facing truly non-perturbative aspects 
of renormalization.

\section{The Solution} \label{sect4}

Faced with such a problem, 
the only way to eliminate this cutoff dependence
is to modify our leading order calculation, that is,
to change our
accounting of the
higher-energy behavior of the theory
through the addition of at least one new interaction.

In order to do so,
we can follow essentially two routes.
One is to revise our systematic expansion in the two-body system.
The task in this case is to enlarge the EFT in order to
incorporate higher-energy degrees of freedom that have so far
been treated only in something akin to a multipole expansion.
Generically, the next mass scale above $1/a_2$ is 
$1/r_2$, and incorporating momenta $Q\sim 1/r_2$ means
that amplitudes certainly can no longer
be treated in a $Qr_2$ expansion.
With these new elements, a new small parameter has to
be identified. 
The resulting, new leading-order two-body amplitude might very well exhibit
sufficiently nice ultraviolet behavior to guarantee the disappearance
of the pronounced cutoff dependence found 
above in the three-body amplitudes.

While this is a perfectly legitimate way to proceed,
it is in most cases not a simple task. It demands a qualitative new
understanding of the shorter-range dynamics of the system under study.
It is also specific to that system since the dynamics
at $Q\sim 1/r_2$ will have to be treated in more detail;
its consequences will not be common to all systems with a large $a_2$.
Moreover, the modifications of the leading-order
two-body amplitude, if any, might not resolve the leakage problem in
the three-body system.
We follow here the other possible route. 
By construction, this approach preserves the expansion of the two-body
amplitude; as such it is simple,
model independent, and designed to function.

The observed cutoff dependence comes from the behavior of the amplitude in
the ultraviolet region, where the EFT Lagrangian, Eq. (\ref{lagt}), is not
to be trusted.
When the low-energy expansion is perturbative the cutoff-dependent
contribution  from high loop momenta can be expanded in powers of
the low external momenta and cancelled by terms in the Lagrangian,
since those also give contributions
analytic in the external momenta. Thus all uncertainty coming
from the high momentum behavior of the theory is
parametrized by the coefficients of a few local counterterms.
The present case is complicated by the fact that the cutoff
dependence of the amplitude is non-analytic around $p=0$.
This, however, does not mean that the renormalization program
in this low-energy EFT is doomed: a three-body force term of 
sufficient strength
contributes not only at tree level, but also in loops
dressed by any number
of two-particle interactions. 
We want to show that 
the bare three-body force 
coefficient 
$h(\Lambda)$ can be combined with the 
the problematic parameter $p_\star(\Lambda)$ in order to produce 
an amplitude $K(k, p)$ which is cutoff independent, at least around 
$p=k=0$. 
Because the cutoff dependence generated by the two-body amplitude 
is a somewhat complicated oscillation, we expect a similar
behavior of the $h(\Lambda)$ that ensures cutoff independence. 
%three-body scattering length $a_3$. 
In particular, $h(\Lambda)$ must be such that it vanishes for 
``critical'' cutoffs $\Lambda_n$ that belong to the family
that gives the desired $a_3$.
For any other cutoff, $h(\Lambda)$ must be non-vanishing.

So, let us now consider the effects of a non-zero $h(\Lambda)$. 
It is convenient to rewrite the
three-body force as $h(\Lambda)=2mg^2 H(\Lambda)/\Lambda^2$.
In the remainder of this paper we will refer to 
$H(\Lambda)$ as the bare three-body force.
$H(\Lambda)$ has to be at least big enough so as to give a non-negligible
contribution in the $p\sim k\sim \Lambda$ region. 
This means that the dimensionless quantity
$H(\Lambda)$ has to be at least of $O(1)$. 
We will look for a solution 
initially assuming a ``minimal'' {\it ansatz} $H(\L)\sim 1$;
we subsequently show via an unbiased
numerical analysis that this assumption is justified.

Such a  three-body force has the feature that its
contribution to the inhomogeneous term is small compared
to the contribution from the two-body interaction, as it is at most 
$p/\Lambda$ of the latter.
When $p\gg 1/a_2$ (but $k\sim 1/a_2$), the amplitude satisfies a new 
approximate equation
\begin{equation}
K(k,p)= \frac{4\lambda}{\sqrt{3}\pi}
                   \int_0^\Lambda \: \frac{dq}{q}\ 
                   \left[{\rm ln} \left(\frac{q^2+pq+p^2}{q^2-pq+p^2}\right)
                   +2H(\Lambda)\frac{pq}{\Lambda^2}\right] K(k,q).
\label{newasphieq} 
\end{equation}
\noindent
For $p\sim \Lambda$ the
term proportional to $H(\Lambda)$ becomes important and
the solution $K(k,p\sim \L)$ has a complicated form.
In the range $1/a_2\ll p\ll \Lambda$, on the other hand, 
the three-body force term
is suppressed by $p/\Lambda$ compared to the logarithm
and can be disregarded. Consequently, the
behavior
(\ref{assol}) is still correct in the intermediate region.
The effect of a
finite value of $H$ can be at most to change the values of the amplitude
$C$ and the phase $\delta$, which become dependent on $H$.
As shown in Fig. \ref{fig5}, this feature
is confirmed by
numerical solutions:
while different values of the three-body force $H$ 
(at a fixed $\L$) preserve the form of the
solution, the phase (and amplitude)  are changed.
If $H$ is chosen to be a function of $\Lambda$
in such a way as to cancel the explicit $\Lambda$ dependence,
we can make the solution of Eq. (\ref{aeq})
cutoff independent for all $p\ll \Lambda$.
In particular, the
scattering amplitude that is determined by the on-shell value
$K(k,k)$ with $k\sim 1/a_2$ will be cutoff independent as well.
For this to be possible $C$ and $\delta$ must depend on
the same combination of $\Lambda$ and $H$.
Thus $H(\Lambda)$ must be chosen such that
\begin{equation}
\label{Lbar}
-s_0 \ln \Lambda + \delta(H(\Lambda))\equiv -s_0 \ln \Lambda_\star ,
\end{equation}
\noindent
where $\Lambda_\star$ is a (cutoff independent)
parameter fixed either by experiment or by matching
with a microscopic model.
Eq. (\ref{Lbar}) simply means that $p_\star(h(\Lambda),\Lambda)= \Lambda_\star$
is cutoff independent, and 
the $1/a_2\ll p\ll \Lambda$ solution is
\begin{equation}
K(k,\Lambda\gg p\gg 1/a_2)= 
       -\frac{\gamma}{\cos(s_0 \ln (\L_\star a_2)+\epsilon)} \: 
        \cos \left(s_0 {\rm ln} \frac{p}{\Lambda_\star}\right). 
\label{newassol}
\end{equation}
\noindent
Matching with the $p\sim 1/a_2$ solution should then determine
the scattering length $a_3=a_3(\Lambda_\star)$ and the low-energy dependence
of the amplitude. 

\begin{figure}[tb]
\begin{center}
\epsfxsize=8cm
\centerline{\epsffile{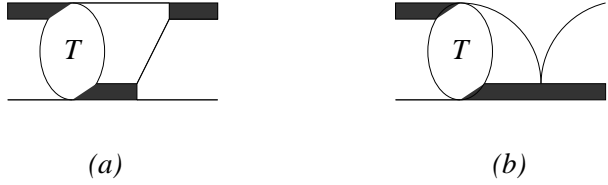}}
\end{center}
\caption{Inhomogeneous terms of the integral equation (\ref{aeq}): 
(a) two-body and (b) three-body kernels.}
\label{fig7}
\end{figure}

For $p\sim 1/a_2$, Eq. (\ref{aeq}) becomes 
\begin{eqnarray}
\frac{3}{4}  
\frac{K(k,p)}{\left(\frac{1}{a_2}+\sqrt{\frac{3p^2}{4}-mE}\right)}
& = &\frac{1}{pk}{\rm ln}\left(\frac{p^2+pk+k^2-ME}{p^2-pk+k^2-ME }\right)
\nonumber \\& & 
+\frac{2\lambda}{\pi p}\int_0^\mu \: dq\ 
{\rm ln}
    \left(\frac{p^2+pq+q^2-ME}
               {p^2-pq+q^2-ME}\right) \frac{q K(k,q)}{q^2-k^2}
\nonumber \\
& & +\frac{4\lambda}{\pi} \int_\mu^\L \: dq 
    \left[\frac{1}{q^2}
    +\frac{H(\L)}{\L^2}\right] K(k,q), \label{lowphieq} 
\end{eqnarray}
\noindent
where $\mu$ is an arbitrary scale such that $\mu\ll \L$,
and we have dropped terms which are smaller by powers of $p/\L$
or $\mu/\L$.
We want $K(k, p\sim 1/a_2)$ cutoff independent and determined 
completely by $\L_\star$. Then all
cutoff dependence contained  
in the $\mu$-to-$\L$ integral ---in the integration limit, implicit
in $K(k,p \gg 1/a_2)$, and in $H(\L)$--- has to combine 
to produce a $\L$-independent result. 
Achieving this means that
$K(k,p\sim 1/a_2)$ will depend only on $\L_\star$ not
just at $p=0$ but also for a range
of $p$'s, because the high-momentum contributions from particle exchange
and from the three-body force have the same momentum dependence. 
The high-momentum modes in Fig. \ref{fig7}(a) can be absorbed in 
Fig. \ref{fig7}(b).

We can now obtain an approximate expression for $H(\L)$.
For this purpose, let us consider Eq. (\ref{lowphieq}) 
for two different values of the cutoff
$\L$ and $\L'>\L$, whose solutions we denote by $K(k,p)$ and
$K'(k,p)$, respectively.
The equation for
$K'(k,p)$ will have the same form as the one for $K(k,p)$ except for
some extra terms:
\begin{equation}
\frac{4\lambda}{\pi}
\left\{ \int_\L^{\L'} dq \left[\frac{1}{q^2}
    +\frac{H(\L)}{\L^2}\right] K'(k,q)
+ \left( \frac{H(\L')}{\L'^2}-\frac{H(\L)}{\L^2} \right)
         \int_\mu^{\Lambda} dq \: K'(k,q)
\right\}.
\label{diff}
\end{equation}
\noindent 
Assuming that $K'(k,p)$ has the same phase $\cos(s_0 \ln(p/\L_\star))$
as $K(k,p)$ even
for $p\sim \L'$, Eq. (\ref{diff})
becomes
\begin{eqnarray}
\frac{4\lambda}{\pi \sqrt{1+s_0^2}} & &
\left\{ \right.  
  \frac{1}{\L'} \left[\sin(s_0\ln({\L'}/{\L_\star})-{\rm arctg}(1/s_0)) 
 + H(\L') \sin(s_0\ln({\L'}/{\L_\star})+{\rm arctg}(1/s_0))\right]
      \nonumber \\ 
& & \left. \: -(\L' \rightarrow \L)\right\}, 
\label{cond}
\end{eqnarray}
\noindent 
plus terms terms that are smaller by $\mu/\L$.
We can make the terms in (\ref{cond})
vanish if 
\begin{equation}
\label{h}
H(\L)= -\frac{\sin(s_0\ln(\frac{\L}{\L_\star})-
                   {\rm arctg}(\frac{1}{s_0}))}
                 {\sin(s_0 \ln(\frac{\L}{\L_\star})+
                   {\rm arctg}(\frac{1}{s_0}))}.
\end{equation}
\noindent
Note that $H(\L)$ is periodic in $\L$: $H(\L_n)=H(\L)$
for $\L_n= \L \exp(n\pi/s_0)\simeq \L (22.7)^n$, $n$ an integer.
In particular, the three-body force (\ref{h}) vanishes at the critical 
cutoffs $\L_n= \L_0 \exp(n\pi/s_0)$
with $\L_0= \L_\star \exp(\arctan (1/s_0))$, as anticipated. 
Clearly, $H$ is periodic in $\L_\star$ as well.

Since with such $H(\L)$ the inhomogeneous equation
for $K'(k,p)$ is nearly the same as the equation for $K(k,p)$,
it follows that $K'(k,p)$ has also the same
amplitude as $K(k,p)$ in the intermediate region.
That is, $H(\L)$ chosen like Eq. (\ref{h}) exactly compensates
any change in cutoff, so that $K'(k,p)=K(k,p)$
for all values $p\ll \L$ (up to terms suppressed
by $p/\L$).
As a consequence, the on-shell $K$-matrix
$K(k,k)$ 
will be $\L$ independent as long as $k\ll \L$.

\begin{figure}[t]
\begin{center}
\epsfxsize=8cm
\centerline{\epsffile{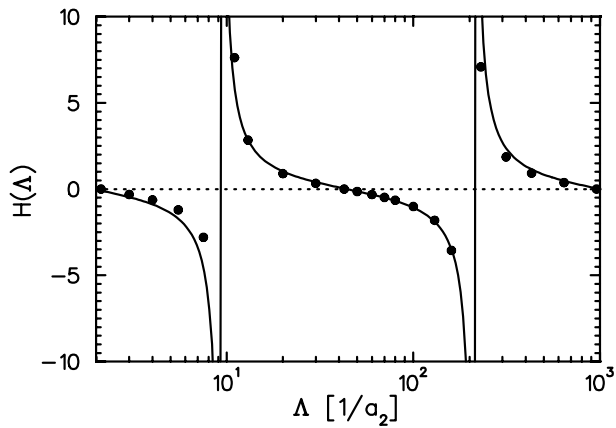}}
\end{center}
\caption{Three-body force $H$ as a function of the cutoff $\Lambda$
for $a_3=1.56 a_2$:
numerical solution (dots) and Eq.(\ref{h})
with $\L_\star=19.5 a_2^{-1}$ (solid line).}
\label{fig8}
\end{figure}

In order to check these arguments we have numerically solved Eq. (\ref{aeq})
for the amplitude $K(k,p)$ with a nonvanishing $H(\L)$, in
the case $\lambda=1$. 
We have fixed the scattering length $a_3= -K(0, 0)$, and
for several cutoffs we have determined the three-body force
that is necessary to keep  $a_3$ 
unchanged.
With such a procedure, we have obtained $H(\L)$ for 
different $\L$'s. In Fig. \ref{fig8} we plot the corresponding
numerical values for the case $a_3=1.56 a_2$, together
with our approximate analytical expression (\ref{h})
with $\L_\star=19.5 a_2^{-1}$. The remarkable agreement confirms that 
the solution for $H(\L)$ is very close to the minimal one, all the way
up to very high values of $\L$.
Note that the three-body force vanishes
at the critical cutoffs in the family generated by
$\L_0= \L_\star \exp(\arctan (1/s_0))=42.6 a_2^{-1}$.\footnote{
Note that the critical cutoffs in Fig. \ref{fig6} were obtained for a 
different $a_3/a_2$.}

\begin{figure}[t]
\begin{center}
\epsfxsize=8cm
\centerline{\epsffile{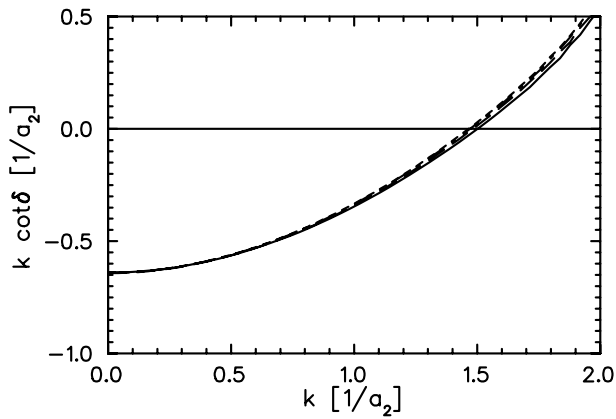}}
\end{center}
\caption{Energy dependence for $\Lambda_\star=19.5 a_2^{-1}$: 
$k\cot\delta$ as function of $k$ for different
cutoffs ($\Lambda=42.6,\,100.0,\,230.0,\,959.0\times a_2^{-1}$).}
\label{fig9}
\end{figure}

\begin{figure}[t]
\begin{center}
\epsfxsize=8cm
\centerline{\epsffile{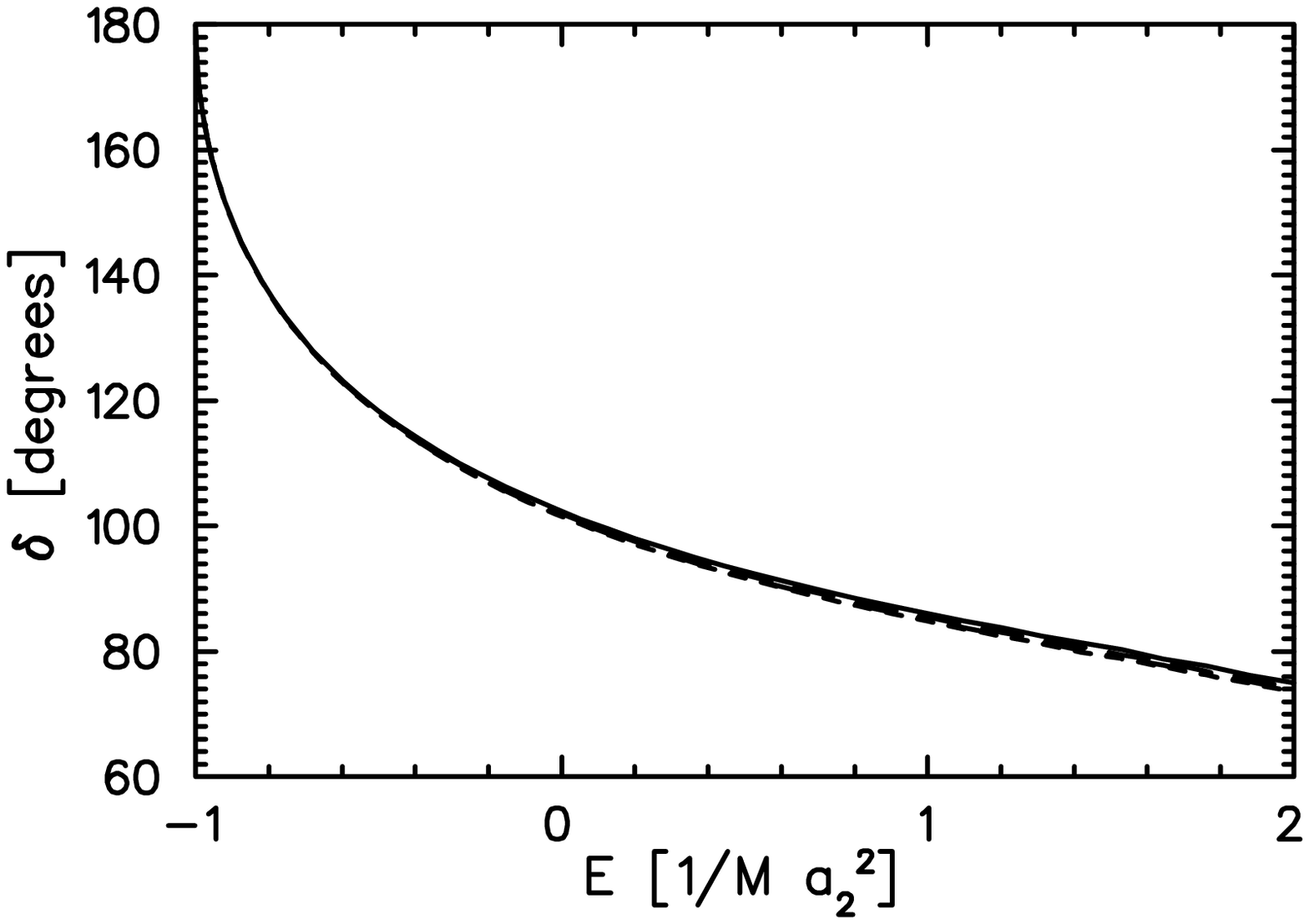}}
\end{center}
\caption{$S$-wave phase shifts $\delta$ as function of energy $E$, as
predicted by the EFT when $a_3=1.56 a_2$ ($\Lambda=42.6,\,100.0,\,230.0,
\,959.0\times a_2^{-1}$).}
\label{fig10}
\end{figure}

In Fig. \ref{fig9}  we show results for 
the corresponding $K(k,k)^{-1}=k \cot\delta$,
where $\delta$ is the $S$-wave phase shift for particle/bound-state
scattering,
for different cutoffs
in the case $a_3=1.56 a_2$.
Comparison with Fig. \ref{fig5} shows that by introduction of
the three-body force we have succeeded in removing
the cutoff dependence of the low-$k$ region,
thus generalizing Fig. \ref{fig6} to cutoffs other than critical. 
As argued above, $k \cot\delta$ is insensitive to
$\Lambda$ as long as $k\ll \Lambda$.
The effective range, for example, is predicted as $r_3= 0.57 a_2$.
In Fig. \ref{fig10} we plot the $S$-wave phase shifts directly.

\begin{figure}[t]
\begin{center}
\epsfxsize=8cm
\centerline{\epsffile{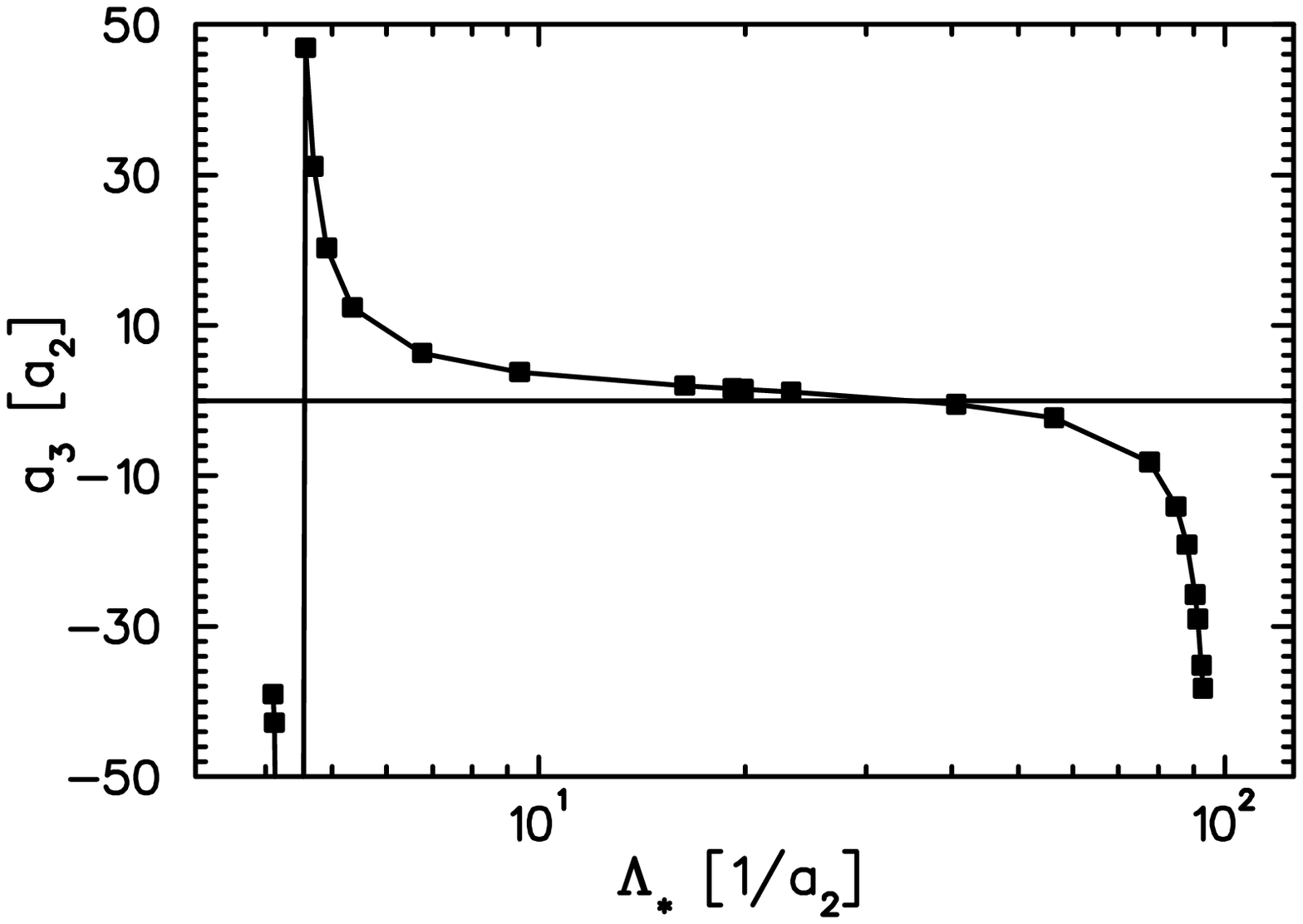}}
\end{center}
\caption{Particle/bound-state scattering length $a_3$ as function
of the three-body force parameter $\Lambda_\star$.}
\label{fig11}
\end{figure}

Similar results can be obtained for other values of $a_3$.
By repeating this procedure we can in fact obtain $a_3=a_3(\Lambda_\star)$
or {\it vice-versa}. 
This is shown in Fig. \ref{fig11}. 
The periodicity of the three-body force with $\Lambda_\star$
is apparent.
If the underlying theory is known and can be solved,
then $\Lambda_\star$ can be determined in terms of underlying parameters,
and $a_3$ can also be predicted.
Otherwise, we can use $a_3$ or any other low-energy three-body datum
to determine $\Lambda_\star$ empirically.
In either case, once the two-body scattering length $a_2$ is fixed,
the low-energy particle/bound-state scattering
amplitude is completely determined by the mass scale $\Lambda_\star$
contained in the three-body force $h$.

\section{Bound State} \label{sect5}

We can extend the preceding analysis to the three-body bound-state problem.
In this case the relevant equation to be solved is Eq. (\ref{aeq})
without inhomogeneous terms. The latter
did not play an important role in our ultraviolet arguments,
so the same arguments hold for any bound
state with binding energy $B_3$ comparable with $1/ma_2^2$,
{\it i.e.}, for any bound state with
the dimensionless binding energy $b_3= ma_2^2 B_3= O(1)$.
Such bound states are shallow, 
having a size $\sim 1/\sqrt{m B_3}= a_2/ \sqrt{b_3}$
comparable to $a_2$,
and thus should be within range of the EFT.
In principle, all bound states with size larger than
$\sim r_2$ should be amenable to this EFT description.
Their properties will be determined in first approximation
by only $a_2$ and $a_3$, or equivalently $C_0$ and $\Lambda_\star$,
while more precise information can be obtained in an
expansion in $\sqrt{b_3}r_2/a_2$.

\begin{figure}[t]
\begin{center}
\epsfxsize=8cm
\centerline{\epsffile{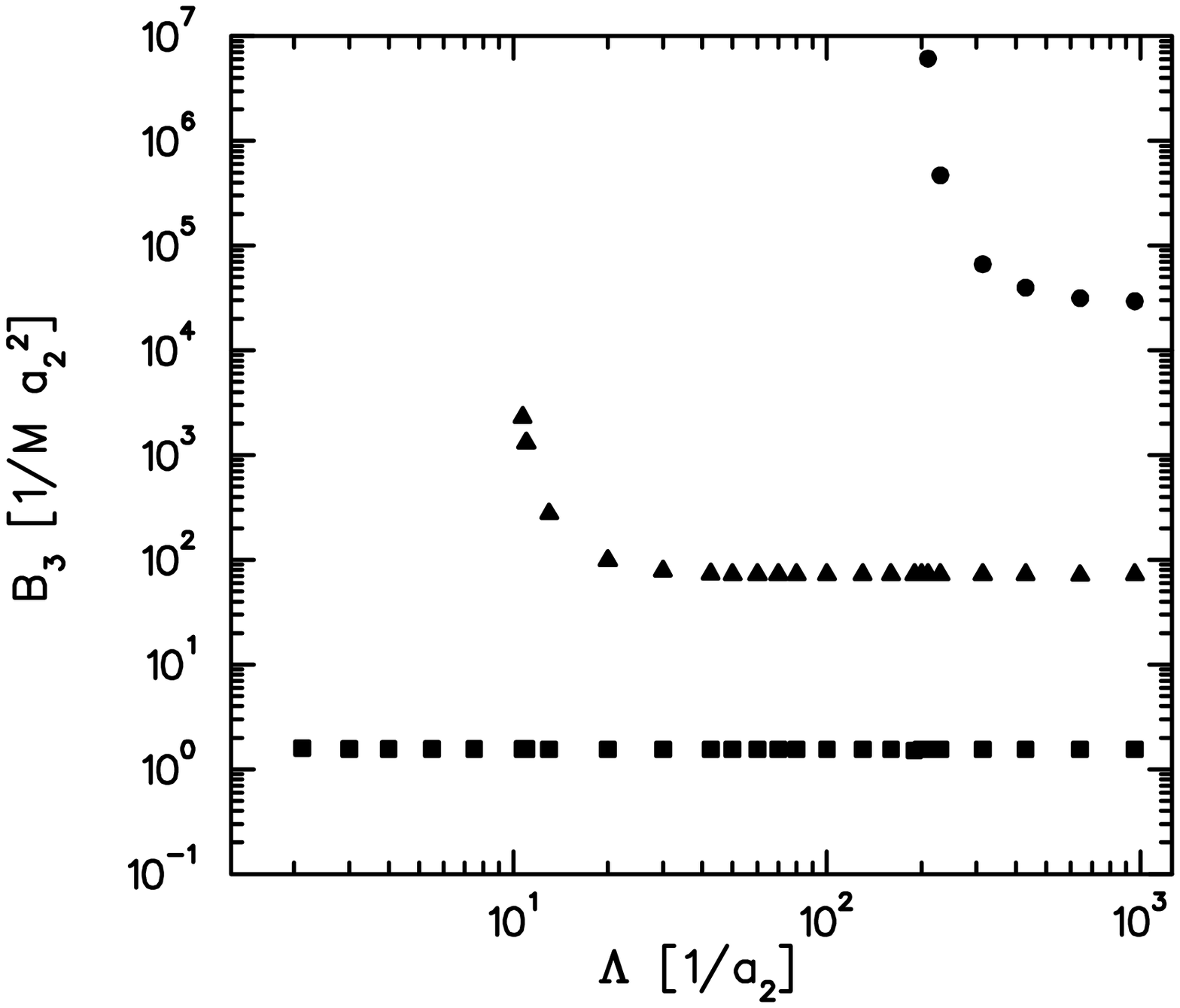}}
\end{center}
\caption{Three-body binding energies $B_3$ as functions of the cutoff 
$\Lambda$ for $\Lambda_\star=19.5 a_2^{-1}$.}
\label{fig12}
\end{figure}

In Fig. \ref{fig12} we plot binding energies 
for a range of cutoffs, with the three-body force adjusted to
give a fixed scattering length $a_3=1.56 a_2$.
(With an appropriate $a_3$,
our results reduce to those of Ref. \cite{kharchenko}
when the cutoff is at one of the critical values $\Lambda_n$.)
As we can see, for this value of $\L_\star$
there exists a shallow bound state at $b_3 \simeq 2$
whose binding energy is independent of the cutoff.
This bound state has a size $\sim 0.7 a_2$ and can thus be 
studied within the EFT.
As we increase the cutoff, deeper bound states appear
at $\Lambda_n= \Lambda_0 \exp(n\pi/s_0)$ with $n$ an integer and 
$\Lambda_0\simeq 10$,
so that for $\Lambda_{n-1}\le \Lambda\le \Lambda_n$ 
there are $n+1$ bound states.
The next-to-shallowest bound state has $b_3 \simeq 70$
and thus a size $\sim 0.1 a_2$. 
If the underlying theory is such that 
$r_2 \saprox a_2/10$, a cutoff $\Lambda\gaprox 1/r_2$ allows
us to deal with the next-to-shallowest bound state.
The next deeper bound state
has $b_3 \simeq 4 \times 10^4$ and thus a size $\sim 0.01 a_2$,
and so on. These most likely lie outside the EFT.

\begin{figure}[tb]
\begin{center}
\epsfxsize=8cm
\centerline{\epsffile{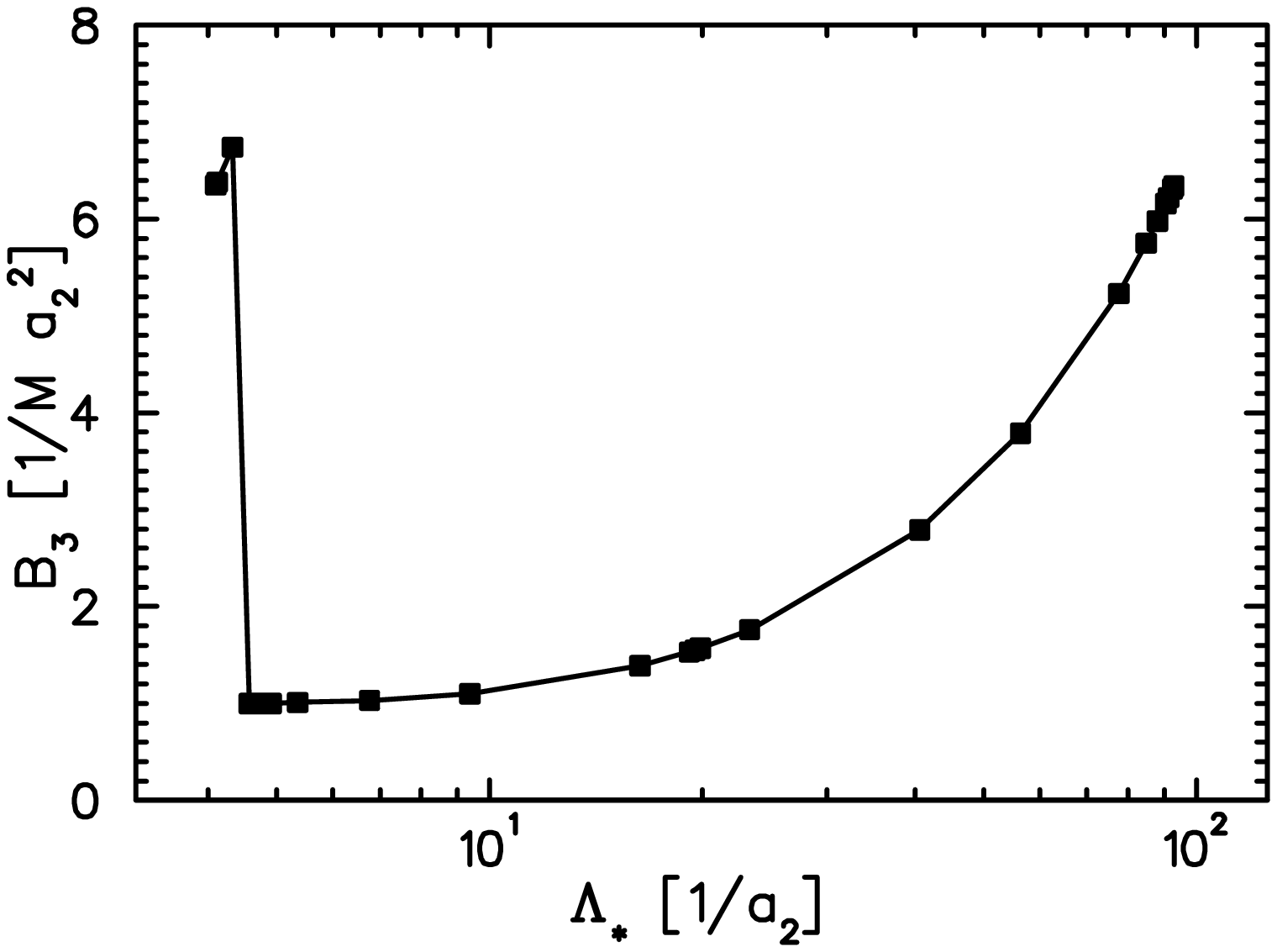}}
\end{center}
\caption{Binding energy $B_3$ of the shallowest bound state
as function of the parameter $\Lambda_\star$.}
\label{fig13}
\end{figure}

Properties of the
shallowest bound state, if within the EFT, are model independent
and thus the most interesting.
For a fixed $a_2$, they are determined by $\Lambda_\star$.
In Fig. \ref{fig13} we plot the binding energy $B_3$
as function of $\Lambda_\star$.
This should be compared with the behavior of $a_3=a_3(\Lambda_\star)$
shown in Fig. \ref{fig11}.
It is clear that at $\L_\star\sim 4 a_2^{-1}$ a bound state appears
at zero energy. As $\L_\star$ increases, this state gets progressively
more bound until at  $\L_\star\sim 22.7 \times 4 a_2^{-1}$ a new 
shallowest bound state
appears; the picture repeats indefinitely.
Whenever a bound state is close to zero energy,
$a_3$ is large in magnitude, negative if the bound state is virtual
and positive if real. This can be seen in Fig. \ref{fig11}.

\begin{figure}[tb]
\begin{center}
\epsfxsize=8cm
\centerline{\epsffile{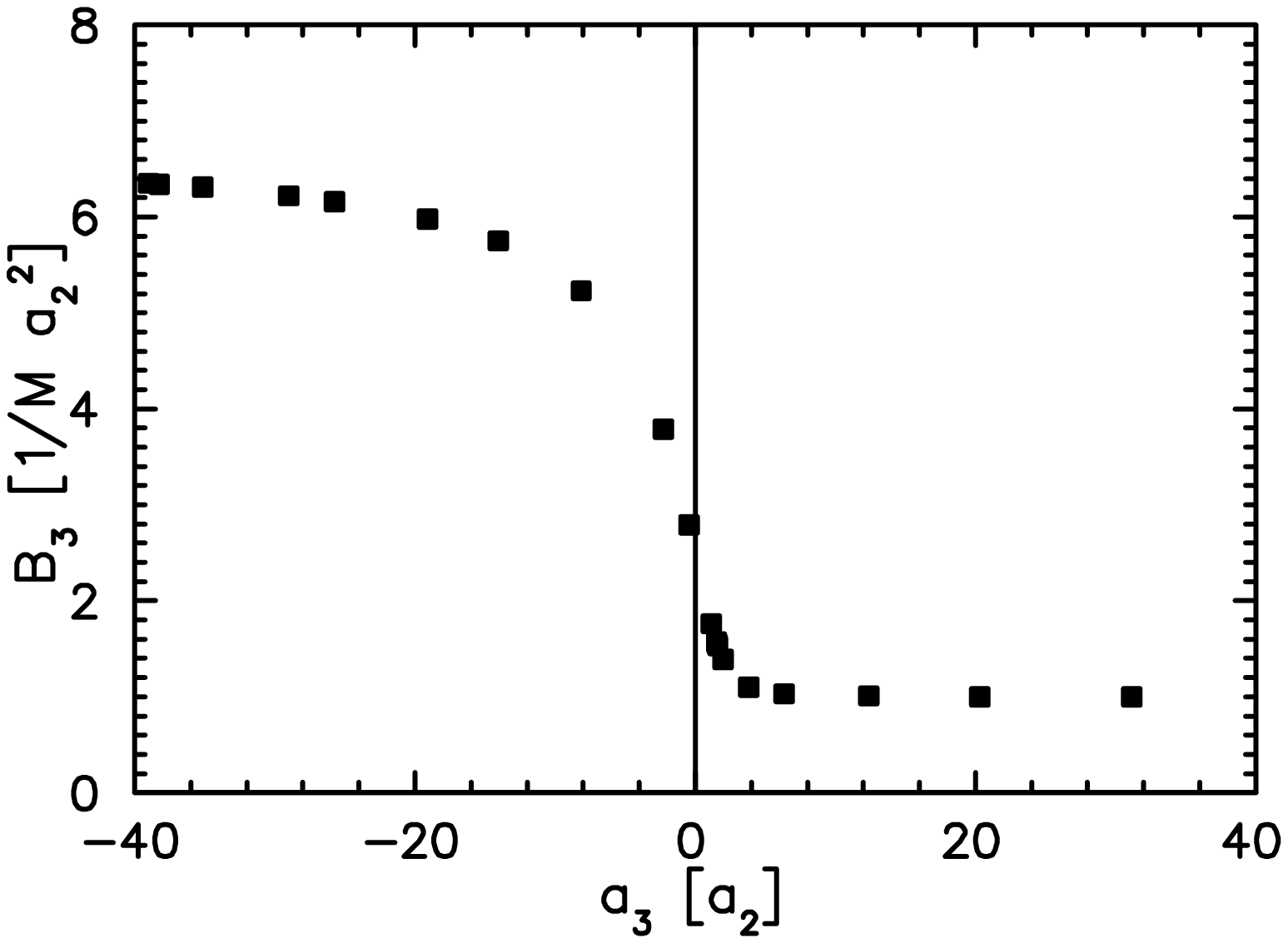}}
\end{center}
\caption{Phillips line: binding energy $B_3$ of the shallowest bound
state as function of the particle/bound-state
scattering length $a_3$.}
\label{fig14}
\end{figure}

Eliminating $\Lambda_\star$, we obtain an universal curve $B_3=B_3(a_3)$,
as in Fig. \ref{fig14}. 
Qualitative features of this curve can be understood
from Figs. \ref{fig11} and \ref{fig13}.
For $a_3$ large in magnitude, small variations of $\Lambda_\star$
lead to large variations of $a_3$ but small variations of $B_3$,
so the curve flattens out at both ends.
For $a_3$ large and positive, there is a shallow real bound state;
for $a_3$ large and negative, there is a shallow virtual bound state
but the curve tracks the deeper real bound state.
For $a_3 \sim a_2$, $a_3$ and $B_3$ vary with $\Lambda_\star$ at a similar
rate, and the curve interpolates between the two ends.

This curve is known in the three-nucleon case as
the Phillips line \cite{phillips}. It has been derived before in the
context of models for the two-particle potential that
differed in their high-momentum behavior.
Varying among two-particle potential models
one could expect to fill up the $B_3\times a_3$ plane.
In the EFT, the high-momentum behavior of the two-particle potential
is butchered, which causes no trouble in describing two-particle scattering
at low-energies, but ---in the peculiar way described here--- does require
a three-body force. 
It is crucial that the EFT does retain a systematic expansion, so that
in leading order it requires {\it only one} local three-body counterterm,
determined by $\Lambda_\star$.
Varying among two-particle potential models
is thus equivalent to varying the one parameter $\Lambda_\star$
of the three-body force. This spans a single curve $B_3=B_3(a_3)$
in the $B_3\times a_3$ plane. 
Our argument suggests that the Phillips line is a generic phenomenon
and provides a simple explanation
of its origin.

\section{Higher orders} \label{sect6}

The corrections to the calculations in the previous sections
come from operators
not explicitly written in Eq. (\ref{lag}) and are suppressed
by powers of $k R$ or $R/a_2$. The first correction comes from 
terms involving two derivatives and four nucleons fields \cite{1stooge}:
\begin{equation}
- \frac{C_2}{8} (\psi^\dagger\psi \, \psi^\dagger \nabla^2 \psi 
                 +\ldots) \label{c2}.
\end{equation}
\noindent
They account for the effective range term in the effective range expansion
of particle-particle scattering, and one finds
$C_2 = \pi a_2^2 r_2/m$. As it was done before,
(\ref{c2}) can also be generated
by integrating out the $T$ field from Eq. (\ref{lagt}) if the extra term
\begin{equation}
 g_2 T^\dagger (i\partial_0+\frac{\nabla^2}{4m})T\label{g2},
\end{equation}
\noindent
where $g_2/\Delta= 4m C_2/C_0$, is added to Eq. (\ref{lagt}). 
Indeed, integration 
over the auxiliary field
$T$ generates, besides the terms shown in Eq. (\ref{lag}), 
also a four-nucleon
term involving two space derivatives or one time derivative. As usual,
a nucleon-field redefinition involving the equations of motion
can be performed to trade the time derivative by two space derivatives.
This field redefinition does not change on-shell amplitudes.
As far as observables are concerned,
adding (\ref{g2}) to Eq. (\ref{lagt}) is thus equivalent 
to adding (\ref{c2}) to Eq. (\ref{lag}).

\begin{figure}[b]
\begin{center}
\epsfxsize=15cm
\centerline{\epsffile{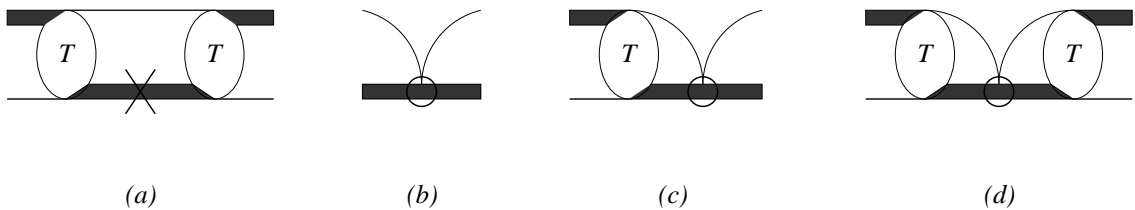}}
\end{center}
\caption{$O(r_2/a_2)$ diagrams: 
(a) one insertion of (\ref{g2}) in the dimeron propagator, denoted by a cross;
(b,c,d) one insertion of a correction to the three-body force $h$,
denoted by a circle.}
\label{fig15}
\end{figure}

The correction proportional to $r_2$ is then given by the diagram
in Fig. \ref{fig15}(a) (plus corresponding wave function renormalization
pieces) and includes one insertion of the kinetic term, Eq. (\ref{g2}). 
The contribution 
of this graph to $t(k,p)$ is
\begin{equation}
\frac{g_2}{2 \pi^2}\int_0^\infty \! dq \, q^2 \, it(q,p) \, 
(i\Delta(3k^2/4m-1/ma_2^2-q^2/2m,\vec{q}))^2 
\left(\frac{3(k^2-q^2)}{4m}-\frac{1}{ma_2^2}\right) it(k,q).\label{r0}
\end{equation}
\noindent
As discussed above, $t(k,p)$ behaves for large $p$ 
as $t(k,p)\sim A(k) \cos (s_0 \log(p/\L_\star))/p$, so the diagram 
is naively linearly divergent. The identity
\begin{equation} \frac{1}{a_2}
\frac{q^2-k^2}{(-\frac{1}{a_2}+\sqrt{\frac{3 (q^2-k^2)}{4}+\frac{1}{a_2^2}})^2}
=\frac{4}{3}\left[ 1+
  \frac{\frac{2}{a_2}(\sqrt{\frac{3 (q^2-k^2)}{4}+\frac{1}{a_2^2}}
                  -\frac{1}{a_2})}
 {\frac{3 (q^2-k^2)}{4}
  -\frac{2}{a_2} 
  (\sqrt{\frac{3 (q^2-k^2)}{4}+\frac{1}{a_2^2}}-\frac{1}{a_2})}\right]
\label{biramagic}
\end{equation}
\noindent
can be used to rewrite Eq. (\ref{r0}) and it
shows that the most divergent piece (the one due to the constant on the 
right hand side of Eq. (\ref{biramagic}))
actually vanishes. This is because in this term the integral can be closed
in the complex plane without circling any pole. The only contribution
comes from the second term in Eq. (\ref{biramagic}) but this term is 
further suppressed in the ultraviolet, resulting in a logarithmically divergent
contribution from the diagram in Fig. \ref{fig15}(a).
This divergent piece has a complicated dependence on the external momentum $k$
since it is proportional to $A^2(k)$ (the dependence on $k$ of the
other terms included in the graph is unimportant in the ultraviolet). 
This is an unusual situation.
In perturbative calculations 
the divergent part is simply a polynomial in the external momenta 
and it can consequently  be absorbed in a finite number of local counterterms
appearing only at tree level. Here the dependence on the external momenta
is more complicated but the counterterm appears in graphs with an arbitrary
number of leading-order interactions. To see this, let us split the
three-body force coefficient $h$ into a leading order piece 
$h^{(0)}$ 
(the same
considered in the previous sections) and a sub-leading piece $h^{(1)}$ that
will be included perturbatively. The inclusion of the sub-leading three-body
force proportional to $h^{(1)}$ generates the three diagrams shown in  
Fig. \ref{fig15}(b,c,d). Simple power counting shows that the graph
in Fig. \ref{fig15}(d) is the most divergent of the three. The important 
point is that the external momentum dependence of the most divergent part of
the graph in Fig. \ref{fig15}(d) is given by $A^2(k)$, again because the 
dependence on the external momentum $k$ of the nucleon propagators in 
Fig. \ref{fig15}(d) can be discarded in the ultraviolet region. Since
the $k$ dependence of the divergent part is the same in
Fig. \ref{fig15}(a) and Fig. \ref{fig15}(d), $h^{(1)}$ can be chosen 
as a function of the cutoff so as to make the sum of these two graphs finite.
As always, the renormalization procedure leaves a finite piece in $h$ 
undetermined that should be fixed through the use of one experimental
input. 
This conclusion seems to agree with  Ref. \cite{efimov2}.
More details of the calculation of the range correction 
as applied to the case of three nucleons in the triton channel 
are left for a future publication.  

\section{Discussion and Conclusion} \label{sect7}

Our results should hold for $^4$He atoms. In fact,
it has recently been established that the two-$^4$He bound state
(``dimer'') is very shallow, with an average size
$\langle r \rangle= 62\pm 10$ \AA \ \cite{dimer},
more than an order of magnitude larger than the
range of the interatomic potential. 
The low-energy two-$^4$He system should then be describable 
by contact two-body forces. In leading order,
the measured size translates into a scattering length
$a_2=124$ \AA, which determines 
the strength of the contact interaction. 
Unfortunately, although the three-$^4$He (``trimer'') has been observed
\cite{trimer}, there seems to be no low-energy
information on its properties nor on $^4$He/dimer scattering.
At least one three-body datum is needed 
to determine $\Lambda_\star$ and use the EFT to make predictions.

Until such datum becomes available, we can only
illustrate the method by using
a phenomenological $^4$He-$^4$He potential
as a model of a microscopic theory.
We select a potential \cite{heliumpot} 
which is consistent with the recent measurement of the dimer binding energy.
It gives for the two-body system
$a_2=124.7$ \AA \ and $r_2\simeq 7.4$ \AA.
Three-body calculations are much more difficult to perform with such
a phenomenological potential.
An estimate for the $^4$He/dimer scattering length
is $a_3=195$ \AA. Ground and excited bound states have been reported;
estimates for the shallowest bound state place it in the
range $B_3= 1.04-1.7$ mK, while a deeper state 
lies around $B_3= 0.082-0.1173$ K.
There exist a prediction for the low-energy $S$-wave phase shifts,
albeit for a different potential,
but $r_3$ could not be determined.

Using such model we can estimate the range of
validity of the EFT in momentum to be $\sim 1/r_2 \simeq 0.14$ \AA$^{-1}$, 
and the leading
order to give an accuracy of $\sim r_2/a_2 \simeq 0.06$,
or about 10\%,
at momenta $Q\sim 1/a_2$.
Using  this potential's $a_3/a_2=1.56$ we can determine $\Lambda_\star$,
and predict both the energy dependence in $^4$He/dimer scattering and the 
trimer binding energy. 
In fact, in Figs. \ref{fig9}, \ref{fig10}, and \ref{fig12}, 
we have used exactly this value of $a_3$.
As a consequence, Fig. \ref{fig9} represents our lowest-order prediction
for $k \cot \delta$ for atom/dimer scattering, from which we can 
for the first time extract
an effective range $r_3= 71$ \AA.
Fig. \ref{fig10} displays the $S$-wave phase shifts themselves,
and is in qualitative agreement with the result of Ref. \cite{heliumpot}
for a similar potential,
being obtained here at a negligible fraction of computer time.
{}From Fig. \ref{fig12} we predict a bound state at
$B_3=1.2$ mK, which is certainly within the EFT. 
The next-to-shallowest bound state is small enough to be at 
the border of EFT applicability. For a sufficiently
large cutoff, we find $B_3=0.057$ K, but in the best case scenario
corrections from higher orders should be a lot larger than 10\%,
and very important.

Because of the similarity of the integral equations,
our arguments should be relevant for systems of three fermions 
with internal quantum numbers as well. We are in the process of
verifying this for the three-nucleon system in the $J=1/2$ channel,
where we will be able to check our predictions against the
energy dependence of existing phase shift analyses of neutron-deuteron
scattering and
against the triton binding energy \cite{more3stooges}.
We also plan to extend our calculations to the four-and-more-body system
and search for the existence of other leading few-body forces.

These results will hold
in an EFT where the pion has been integrated out,
which is valid for $Q\saprox m_\pi$.
In an exciting new development,
it has been argued in Ref. \cite{3musketeers} that 
there is a region of momenta above $m_\pi$ where,
although pions have to be kept explicitly, their effects 
are sub-leading. The leading two-body operators are thus
the same as in the ``pionless'' theory.
A number of examples seem to corroborate this
picture \cite{ocordaumdospuxasacos...}.
In this case, our leading-order results will be valid 
in this ``pionful'' theory as well,
suggesting
that our bound-state calculation will provide a reasonable
estimate of the triton binding energy.

In conclusion,
we have provided analytical and numerical evidence that
renormalization of the three-body problem with short-range forces 
requires in general
the presence in leading order of a one-parameter contact three-body force.
This frames results obtained
earlier with particular models within a larger,
model-independent picture.
It opens up the possibility of applying the EFT method to a large class
of systems of three or more particles with short-range forces.

\vspace{2cm}
{\large \bf Acknowledgments}

We thank V. Efimov, H. M{\"u}ller, and D. Kaplan
for helpful discussions.
HWH and UvK acknowledge the hospitality of the Nuclear Theory Group and
the INT in Seattle, where part of this work was carried out.
This research was supported in part by the U.S. Department of Energy
grants DOE-ER-40561 and DE-FG03-97ER41014, 
the Natural Science and Engineering Research Council of Canada,
and the U.S. National Science Foundation grant PHY94-20470.

\end{document}